\documentclass[a4paper,12pt]{article}
\usepackage[dvips]{graphicx}

\title{Dependence of the Brillouin precursor form on the initial
signal rise time}
\author{Adam Ciarkowski \\
\normalsize Institute of Fundamental Technological Research\\
\normalsize Polish Academy of Sciences}
\date{}

\def\o{\omega}
\def\d{\delta}
\def\t{\theta}
\def\w{\tilde}
\def\r{\rho}
\def\bet{\mathcal{B}}
\def\b{\beta}
\newcommand{\la}{\lambda}
\def\g{\gamma}

\begin{document}

\maketitle

\begin{abstract}
Propagation of a Brillouin precursor in a Lorentz dispersive
me\-dium is considered. The precursor is excited by a sine
modulated initial signal, with its envelope described by a
hyperbolic tangent function. The purpose of the paper is to show
how the rate of growth of the initial signal affects the form of
the Brillouin precursor. Uniform asymptotic approach, pertinent to
coalescing saddle points, is applied in the analysis. The results
are illustrated with numerical examples.

\vspace{3ex} \noindent\textit{Key words:} Lorentz medium,
dispersive propagation, Brillouin precursor, uniform asymptotic
expansions

\end{abstract}

\section{Introduction}
It is well known that if very fast, rapidly oscillating signals
propagate in a real medium, they undergo the dispersion
phenomenon. Various frequency components of a signal propagate
with different phase velocities, and they are differently dumped.
As a result, the shape of the signal is distorted during
propagation the signal in the medium. Naturally, this phenomenon
is practically important only at very short times and very high
frequencies (of the order of $10^{12}$ Hz and above in the assumed
model). In now classical works Sommerfeld \cite{so;14} and
Brillouin \cite{br;14,br;60} have shown that in the Lorentz model
of a dispersive medium, apart of the main signal two small
precursors are formed. In the asymptotic description of the total
field these precursors are interpreted as contributions to the
field resulting from two different pairs of saddle points. For the
Sommerfeld precursor pertinent simple saddle points vary outside
some disc in a complex frequency plane. As the space-time
coordinate $\t$, to be defined later, takes the initial value
equal unity, those points merge at infinity to form one saddle
point of infinite order. As $\t$ grows up to infinity, they
separate into two simple saddle points that move symmetrically
with respect to the imaginary axis towards corresponding branch
points located in the left and the right half-plane, respectively.
In the case of Brillouin precursor, two other simple saddle points
vary inside a smaller disc. As the coordinate $\t$ grows from
unity, they move toward each other along the imaginary axis,
coalesce into one saddle point of the second order on the axis,
and then again split into simple saddle points that depart from
the axis and move, symmetrically with respect to this axis,
towards corresponding branch points in the left and the right
half-plane, respectively. The location of the saddle points
affects local oscillations and dumping of the precursor. It
depends on the space-time coordinate $\t$ and is governed by the
saddle point equation.

In this paper we confine our attention to the Brillouin precursor,
also called a second precursor (as opposed to the first,
Sommerfeld precursor). Fundamental work on this precursor is due
to Brillouin \cite{br;14,br;60}. Because of limitations of
asymptotic methods then available (now referred to as non-uniform
methods), Brillouin could not correctly describe the precursor's
dynamics for values of $\t$ corresponding to the coalescence of
simple saddle points into one saddle point of a higher order. With
the development of advanced, uniform asymptotic techniques,
complete description of the precursor now got feasible (Kelbert
and Sazonov \cite{ks;96}, and Oughstun and Sherman \cite{os;97}).
In the latter monograph, in addition to the delta function pulse,
the unit step-function modulated signal and the rectangular
modulated signal, the authors also studied an initial signal with
finite rate of growth. In their model, however, the envelope of
the initial signal is described by everywhere smooth function of
time, tending to zero as time goes to minus infinity
(\cite{os;97}, Sec.~4.3.4). In the present paper we consider more
realistic excitation which is caused by an abruptly switched
modulated sine signal, vanishing identically for time $t<0$ and
being non-zero for $t>0$. At $t=0$ the derivative of the signal's
envelope suffers a step discontinuity. As $t$ increases, the
envelope grows with a finite speed, asymptotically tending to its
maximum value. In the following sections we construct uniform
asymptotic representation for the Brillouin precursor resulting
from this sort of excitation, and show how the speed of growth in
the initial signal affects the form of the precursor. We also
illustrate the results with numerical examples.

\section{Formulation of the problem}
We consider a one dimensional electromagnetic problem of
propagation in a Lorentz medium. The medium is characterized by
the frequency-dependent complex index of refraction
\begin{equation}\label{e1}
n(\o) = \left( 1 - \frac{b^2}{\o^2 - \o_0^2 + 2 i \d\o}
\right)^{1/2},
\end{equation}
where $b$ is so called plasma frequency of the medium, $\delta$ is
a damping constant and $\o_0$ is a characteristic frequency.

Any electromagnetic field in the medium satisfies the Maxwell
equations
\begin{eqnarray*}
\nabla\times E({\bf r},t)-\frac{1}{c}\frac{\partial H({\bf
r},t)}{\partial t}=0, \hspace{.5in} \nabla\times H({\bf r},t)-
\frac{1}{c}\frac{\partial E({\bf r},t)}{\partial t}=0,\\
D({\bf r},t)=\int_{-\infty}^t \tilde{\epsilon}(t-\tau) E({\bf
r},\tau) d\tau, \hspace{.4in} B({\bf r},t)=\mu H({\bf
r},t),\hspace{.7in}
\end{eqnarray*}
where $\tilde{\epsilon}(t)$ is a real function and $\mu$ is a real
constant (hereafter assumed to be equal 1). By Fourier
transforming the equations with respect to $t$ and assuming that
the fields depend on one spatial coordinate $z$ only, we obtain
the following equations for transforms of the respected fields
\[
{\bf \hat{z}}\times {\cal H}(z,\o)=-\frac{i\o\epsilon(\o)}{c}{\cal
E}(z,\o), \hspace{.5in} {\bf \hat{z}}\times {\cal
E}(z,\o)=\frac{i\o\mu}{c}{\cal H}(z,\o),
\]
where ${\bf \hat{z}}$ is the unit vector directed along $z$-axis
and $\epsilon(\o)=n^2(\o)/(c^2\mu)$ is the Fourier transform of
$\breve{\epsilon}(t)$. It then follows that ${\bf \hat{z}}$,
${\cal E}$ and ${\cal H}$ are mutually perpendicular. Moreover, if
${\cal E}$ is known then ${\cal H}$ is also known, and vice versa.
It is also true for the electromagnetic field components, which
are the inverse Fourier transforms of ${\cal E}$ and ${\cal H}$.
Therefore, the knowledge of the electric (magnetic) field is
sufficient to determine the full electromagnetic field. To make
the calculations as simple as possible, it is advisable that the
$x$ (or $y$) axis be directed to coincide with the electric or
magnetic field.

Assume that in the plane $z=0$ an electromagnetic signal is turned
on at the moment $t=0$. For $t>0$ it oscillates with a fixed
frequency $\o_c$ and its envelope is described by a hyperbolic
tangent function. Suppose the selected Cartesian component (say
$x$-component) of one of these fields in the plane $z=0$ is given
by
\begin{equation}\label{e2}
A(0,t)=\left\{
\begin{array}{ll}
0 &               t<0 \\
\tanh \b t\sin \o_ct & t\ge 0,
\end{array}
\right.
\end{equation}
The parameter $\b$ determines how fast the envelope of the signal
grows.

This initial electromagnetic disturbance excites a signal $A(z,t)$
outside the plane $z=0$. In what follows we will be interested in
the field propagating in the half-space $z>0$. The problem under
investigation can be classified as a mixed, initial-boundary value
problem for the Maxwell equations.

The exact solution for this specific form of the initial signal
$A(0,t)$ is described by the contour integral \cite{ac;97}
\begin{equation}\label{e3}
A(z,t)=\int_C g(\o) e^{\frac{z}{c}\phi(\o,\t)}\,d\o,
\end{equation}
defined in the complex frequency plane $\o$. Here,
\begin{eqnarray}\label{e4}
\lefteqn{g(\o)=} \nonumber \\
& & \frac{1}{4\pi}
\left[\frac{i}{\b}\bet\left(-\frac{i(\o-\o_c)}{2\b}\right) +
\frac{1}{\o-\o_c}
-\frac{i}{\b}\bet\left(-\frac{i(\o+\o_c)}{2\b}\right) -
\frac{1}{\o+\o_c} \right], \hspace{.8cm}
\end{eqnarray}
the complex phase function $\phi(\o,\t)$ is given by
\begin{equation}\label{e5}
\phi(\o,\t)=i\,\frac{c}{z}\,[\w{k}(\o)z-\o t]=i\o[n(\o)-\t],
\end{equation}
and $\bet(s)$ is the beta function \cite{rg;51} defined via the
psi function as
\begin{equation}\label{e6}
\bet(s) = \frac{1}{2}\left[\psi\left(\frac{s+1}{2}\right) -
\psi\left(\frac{s}{2}\right)\right].
\end{equation}
$\bet(s)$ is a Fourier transform of the initial signal envelope
$\tanh{\b t}$. The dimensionless parameter
\begin{equation}\label{e7}
\t=\frac{c t}{z}
\end{equation}
defines a space-time point $(z,\,t)$ in the field, and $c$ is the
speed of light in vacuum. The contour $C$ is the line
$\o=\o^\prime+ia$, where $a$ is a constant greater than the
abscissa of absolute convergence for the function in square
brackets in (\ref{e4}) and $\o^\prime$ ranges from negative to
positive infinity.

Our goal is twofold. First, we shall seek an asymptotic formula
for the second (Brillouin) precursor that results from the
excitation $A(0,t)$. In other words, we shall find near saddle
points contribution to the uniform asymptotic expansion of the
total field $A(z,t)$. Second, we shall examine how the speed
parameter $\b$ in (\ref{e2}) affects the form of the Brillouin
precursor.

\section{Asymptotic representation for the second precursor}
Our derivation of the asymptotic formula for the Brillouin
precursor is based on the technique developed by Chester et al.\
\cite{cf;57} for two simple saddle points coalescing into one
saddle point of the second order. The technique is also
conveniently described in \cite{fm;73} and \cite{bh;75}.

The locations in the complex $\o$-plane of the saddle points in
(\ref{e3}) are determined from the saddle point equation
\begin{equation}\label{e8}
n(\o) + \o n'(\o) - \t = 0.
\end{equation}
At these points the first derivative %$\partial\phi[\o(\t),\t]/\partial\o$
$\phi^{'}_{\o}(\o,\t)$ of the phase function vanishes. We are
interested in the near saddle points, varying in the domain $|\o|<
\sqrt{\o_0^2+\d^2}$. As $\t$ increases from 1 to a value denoted
by $\t_1$, the near saddle points $\o_1$ and $\o_2$  approach each
other along the imaginary axis from below and from above,
respectively (\cite{os;97}). They coalesce to form a second order
saddle point at $\t=\t_1$. Finally, as $\t$ tends to infinity they
depart from the axis and symmetrically approach the points
$\o_{1,2}=\pm\o_0-i \d$ in the right and in the left complex $\o$
half plane, respectively. If $n(\o)$ is eliminated from (\ref{e8})
then the equation can be represented in the form of an eighth
degree polynomial in $\o$ on its left hand side, and zero on its
right hand side. It does not seem to be possible to solve the
equation exactly. In what follows we shall employ the solution to
(\ref{e8}) which was obtained numerically. Alternatively, a simple
approximate solution found in \cite{ac;02} could be used here at
the expense of accuracy in resulting numerical examples.

The first step in the procedure is to change the integration
variable $t$ in (\ref{e3}) to a new variable $s$, so that the map
$s(\o)$ in some disk $D$ containing the saddle points $\o_\pm$
(but not any other saddle points) is conformal and at the same
time the exponent takes the simplest, polynomial form
\begin{equation}\label{e14}
\phi(\o,\t)=\r+\g^2 s-\frac{s^3}{3}\equiv \tau(s,\t).
\end{equation}
Notice that $\tau(s,\t)$ has two simple saddle points  $s=\pm\g$
that can coalesce into one saddle point $s=0$ of the second order,
corresponding to $\o=\o_s$. From
\begin{equation}\label{e15}
\dot{\o}(s)=\frac{\g^2-s^2}{\phi^{'}_{\o}(\o,\t)}
\end{equation}
we infer that for $s(\o)$ to be conformal, $s=\g$ should
correspond to $\o=\o_1$, and $s=-\g$ should correspond to
$\o=\o_2$. Then,
\begin{equation}\label{e16}
\dot{\o}(\pm\g)=\sqrt{\frac{\mp 2\g}{\phi^{''}(\o_{1,2})}},
\end{equation}
where $\phi^{''}(\o_{1,2})$ is a short notation for
$\phi^{''}(\o_{1,2},\t)$. In case the saddle points $\o=\o_{1,2}$
merge to form one saddle point of the second order $\o=\o_s$, one
has $\phi^{''}(\o_s)=0$, and the relevant formula for $\dot{\o_s}$
is
\begin{equation}\label{e17}
\dot{\o}(0)=\left[\frac{-2}{\phi^{'''}(\o_s)}\right]^{1/3}.
\end{equation}
By using correspondence $\o_{1,2}\leftrightarrow\pm\g$ in
(\ref{e14}) one finds that $\g^3$ and $\r$ are equal to
\begin{equation}\label{e19}
\frac{4\g^3}{3}=\phi(\o_1)-\phi(\o_2) ,
\end{equation}
\begin{equation}\label{e20}
\r=\frac{1}{2}[\phi(\o_1)+\phi(\o_2)].
\end{equation}
The equation (\ref{e19}) for $\g$ has three complex roots. Only
one root corresponds to a regular branch of the transformation
(\ref{e14}) leading to the conformal map $s(\o)$. To find the
proper value of $\g$ we first note from (\ref{e17}) that
$\arg{\dot{\o}}(0)$ can take one of the three values: $\pi/6$,
$5\pi/6$ or $-\pi/2$ corresponding to three different branches of
the transformation (\ref{e14}). It can be readily verified that
for $\t<\t_s$ both $\phi(\o_1)$ and $\phi(\o_2)$ are real valued
and $\phi(\o_1)>0$, $\phi(\o_2)<0$. Then it follows from
(\ref{e19}) that $\g^3>0$. On the other hand, if $\t>\t_s$ then
$\o_1=-\o_2^\ast$.\footnote{The star denotes complex conjugate.}
This implies that in the present case
$\phi(\o_1)=[\phi(\o_2)]^\ast$, and similarly $\phi^{''}(\o_1) =
[\phi^{''}(\o_2)]^\ast$. It is now seen that RHS of (\ref{e14})
equals $-2i\;\hbox{Im}\;\phi(\o_2)$, where
$\hbox{Im}\;\phi(\o_2)<0$. Hence for $\t>\t_s$,
$\arg{\g^3}=\pi/2$. We now take advantage of the fact that
$\dot{\o}(s)$ as given by (\ref{e16}) tends in the limit to
(\ref{e17}) as $s\rightarrow 0$. Because $\arg{\phi(\o_1)}=0$ and
$\arg{\phi(\o_2)}=\pi$ for $\t<\t_s$, and
$\arg{\phi(\o_1)}\rightarrow -\pi/2$ and
$\arg{\phi(\o_2)}\rightarrow \pi/2$ as $\t\rightarrow\t_s^+$, we
conclude that $\arg{\g}=0$ $\t<\t_s$ and $-\pi/2$ for $\t>\t_s$,
i.e.
\begin{equation}\label{e20a}
\g=[\frac{3}{4}|\phi(\o_1)-\phi(\o_2)|]^{1/3}\;e^{i\alpha},
\end{equation}
where $\alpha=0, -\pi/2$ if $\t<\t_s$ or $\t>\t_s$, respectively.

With the new variable of integration the integral (\ref{e3}) can
be written down in the form
\begin{equation}\label{e16a}
A(z,t)= \int_{C_1\cap \hat{D}} G(s,\t) e^{\la\tau(s,\t)}\,ds +
\mathcal{E},
\end{equation}
where $\la=z/c$ and
\begin{equation}\label{e17a}
G(s,\t)=g[\o(s)]\; \dot{\o}(s).
\end{equation}
The contour $C_1$ is an infinite arc in the left $s$ complex
half-plane, symmetrical with respect to the real axis, running
upwards and having rays determined by the angles $-i2\pi/3$ and
$i2\pi/3$ as its asymptotes. The domain $\hat{D}$ is the image of
$D$ under (\ref{e14}). The term $\mathcal{E}$, standing for the
integral of $G(s,\t) \exp{\la\tau(s,\t)}$ defined over the parts
of $C_1$ outside $\hat{D}$, is exponentially smaller than $A(z,t)$
itself.

We now represent $G(s,\t)$ in the canonical form
\begin{equation}\label{e22}
G(s,\t)=c_0+c_1s+(s^2-\g^2)H(s,\t).
\end{equation}
Provided the function $H(s,\t)$ is regular, the last term in
(\ref{e22}) vanishes at the saddle points $s=\pm\g$, and its
contribution to the asymptotic expansion is smaller than that from
the first two terms. Indeed, it can be shown that integration by
parts of the last term leads to an integral of similar form as
(\ref{e16}) multiplied by $\la^{-1}$.

\noindent To determine $c_0$ and $c_1$ we substitute $s=\pm\g$ in
(\ref{e22}) and thus find
\begin{equation}\label{e23}
c_0=\frac{G(\g,\t)+G(-\g,\t)}{2}
\end{equation}
\begin{equation}\label{e24}
c_1=\frac{G(\g,\t)-G(-\g,\t)}{2\g}.
\end{equation}

\noindent By using (\ref{e22}) and (\ref{e14}) in (\ref{e16}), and
extending the integration contour in the resulting integrals to
$C_1$, we find that the leading term of the asymptotic expansion
of $A(z,t)$ as $\la\rightarrow\infty$ is given by
\begin{equation}\label{e30}
  A(z,t)\sim 2\pi i
  e^{\la\r(\t)}\left(\la^{-1/3}c_0(\t)
  \hbox{Ai}[\la^{2/3}\g(\t)^2]+\la^{-2/3}c_1(\t)
  \hbox{Ai}^{\prime}[\la^{2/3}\g(\t)^2]\right).
\end{equation}
It is defined through the Airy function and its derivative, as
given by (\cite{bh;75})
\begin{equation}\label{e29}
 \hbox{Ai}(x)=\frac{1}{2\pi i}\int_{C_1} e^{sx-s^3/3}ds
 \qquad \hbox{Ai}^{\prime}(x)=\frac{1}{2\pi i}\int_{C_1}
 s\,e^{sx-s^3/3}ds.
\end{equation}
Plots of both functions for real $x$ are shown in Fig.~1.

\begin{figure}[ht]
\centering
\includegraphics[width=.7\textwidth]{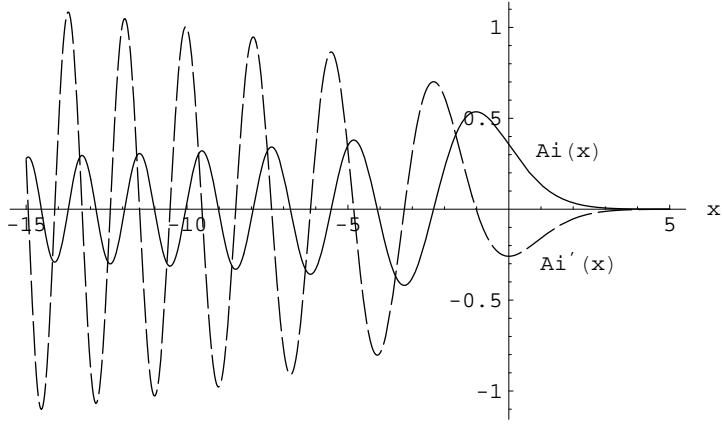}
\caption{\textit{Plots of Ai and Ai' against a real argument x.}}
\end{figure}

The expansion holds for any $\g(\t)$, including $\g=0$. This
special case corresponds to coalescing of the two simple saddle
points $\tau=\pm\g$ into one saddle point of the second order. In
other words the expansion is uniform in $\g$, and hence in $\t$.
It is seen that for $\g=0$, i.e.\ for $\t=\t_s$, the algebraic
order of $A(z,t)$ in $\la$ is $\la^{-1/3}$. This behavior is
characteristic of an integral with a saddle point of the second
order.

For $\g$ well separated from zero the Airy function and its
derivative can be replaced by their asymptotic expansions
(\cite{as;64})
\begin{equation}\label{e30a}
\hbox{Ai}(x)\sim e^{-\frac{2x^{3/2}}{3}} \left[
\frac{1}{2\sqrt{\pi}x^{1/4}}+O(x^{-3/4}) \right],
\end{equation}
\begin{equation}\label{e31}
\hbox{Ai}^{\prime}(x)\sim e^{-\frac{2x^{3/2}}{3}} \left[-
\frac{x^{1/4}}{2\sqrt{\pi}}+O(x^{-3/4}) \right]
\end{equation}
as $x\rightarrow\infty$, and
\begin{eqnarray}\label{e32}
\lefteqn{\hbox{Ai}(x)\sim \frac{1}{\sqrt{\pi}(-x)^{1/4}}\left\{
\sin\left[\frac{2}{3}(-x)^{3/2}+\frac{\pi}{4}\right]
\left(1+O\left[(-x)^{-2}\right]\right)\right.}\\
& & \left.{} +O\left[(-x)^{-3/2}\right]\right\},
\hspace{3in}\nonumber
\end{eqnarray}
\begin{eqnarray}\label{e33}
\lefteqn{\hbox{Ai}^{\prime}(x)\sim
-\frac{(-x)^{1/4}}{\sqrt{\pi}}\left\{
\cos\left[\frac{2}{3}(-x)^{3/2}+\frac{\pi}{4}\right]
\left(1+O\left[(-x)^{-2}\right]\right)\right.}\\
& & \left.{} +O\left[(-x)^{-3/2}\right]\right\}
\hspace{3in}\nonumber
\end{eqnarray}
as $x\rightarrow -\infty$. By using these expansions in
(\ref{e30}) we arrive at the following non-uniform asymptotic
representation of the precursor
\begin{equation}\label{e34}
 A(z,t)\sim e^{\la \phi(\o_2)}
 \left(\frac{-2\pi}{\la \phi^{''}(\o_{2})}\right)^{1/2}
 g(\o_2)
\end{equation}
if $\t<\t_s$, and
\begin{equation}\label{e35}
 A(z,t)\sim e^{\la \phi(\o_1)}
 \left(\frac{-2\pi}{\la \phi^{''}(\o_{1})}\right)^{1/2}
 g(\o_1)+
 e^{\la \phi(\o_2)}
 \left(\frac{-2\pi}{\la \phi^{''}(\o_{2})}\right)^{1/2}
 g(\o_2)
\end{equation}
if $\t>\t_s$.

We see from the above formulas that for $\t$ sufficiently distant
from $\t_s$ (for Brillouin's choice of medium parameters
$\t_s\approx 1.5027$), the representation (\ref{e30}) reduces to a
simple saddle point contribution from $\o_2$ if $\t<\t_s$, and to
a sum of simple saddle point contributions from $\o_1$ and $\o_2$
if $\t>\t_s$. In this manner it is confirmed that the saddle point
$\o_1$ does not contribute when $\t<\t_s$. This is a direct
consequence of the fact that the original contour of integration
in (\ref{e3}) cannot be deformed to a descent path from imaginary
$\o=\o_1$. The algebraic order of $A(z,t)$ in $\la$ is now
$\la^{-1/2}$ because in this case separate simple saddle points
contribute to the expansion. From (\ref{e34}) and (\ref{e35}) it
is also seen that these formulas are non-applicable at $\t=\t_s$
(i.e.\ $\g=0$), where $\phi^{''}(\o_{1,2})=0$. On the other hand
the uniform expansion (\ref{e30}) remains valid for any $\t$ (and
$\g$). In particular it provides a smooth transition between the
cases of small and large $|\g|$.

If $\t>\t_s$, then it can be readily seen that
$g(\o_1)=g^*(\o_2)$, $\phi(\o_1)=\phi^*(\o_2)$, and similarly
$\phi^{''}(\o_1)=\phi^{{''}^*}(\o_2)$. In this case (\ref{e35})
can be written down in a more compact form
\begin{equation}\label{e36}
 A(z,t)\sim 2 \mbox{Re} \left[
 e^{\la \phi(\o_2)}
 \left(\frac{-2\pi}{\la \phi^{''}(\o_{2})}\right)^{1/2}
 g(\o_2)\right].
\end{equation}
Fig.~2 shows the dynamics of the Brillouin precursor in a Lorentz
medium as given by its uniform representation (\ref{e30}) and
non-uniform ones (\ref{e34}) and (\ref{e35}). Throughout this work
the Brillouin's choice of medium parameters
\begin{equation}\label{e36b}
b=\sqrt{20.0}\times 10^{16} s^{-1}, \quad \o_0=4.0\times 10^{16}
s^{-1}, \quad \d=0.28\times 10^{16} s^{-1}
\end{equation}
is assumed.

\begin{figure}[h]
\centering
\includegraphics[width=.7\textwidth]{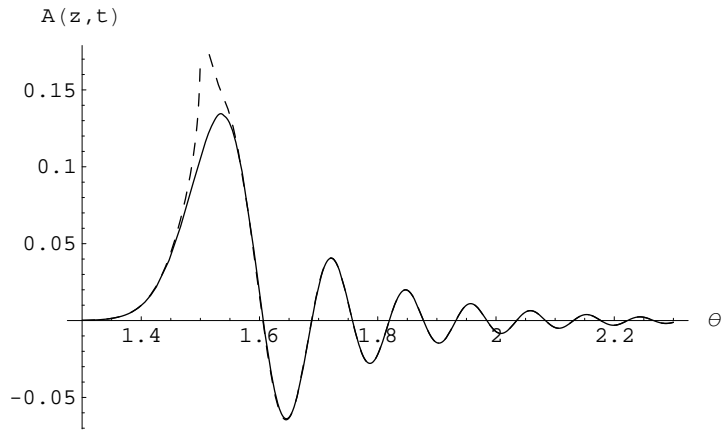}
\caption{\textit{ Uniform (solid line) and non-uniform (dashed
line) representation of the Brillouin precursor in a Lorentz
medium described by Brillouin's parameters. Here, $\b=1.0\times
10^{19} s^{-1}$, $\o_c=2.5\times 10^{16} s^{-1}$ and
$\la=3.0\times 10^{-15} s$.}}
\end{figure}

For $\t<\t_s$ the function $\g(\t)$ takes positive values and the
precursor is described by a monotonically changing function.
Adversely, for $\t>\t_s$ the argument in both functions takes
negative values which leads to oscillatory behavior of the
precursor. This reflects the behavior of both Airy function and
its derivative for positive and negative values of their argument
(see Fig~1).

\section{Dependance of the precursor on the rate parameter $\b$}
\begin{figure}[ht]
\centering
\includegraphics[width=.7\textwidth]{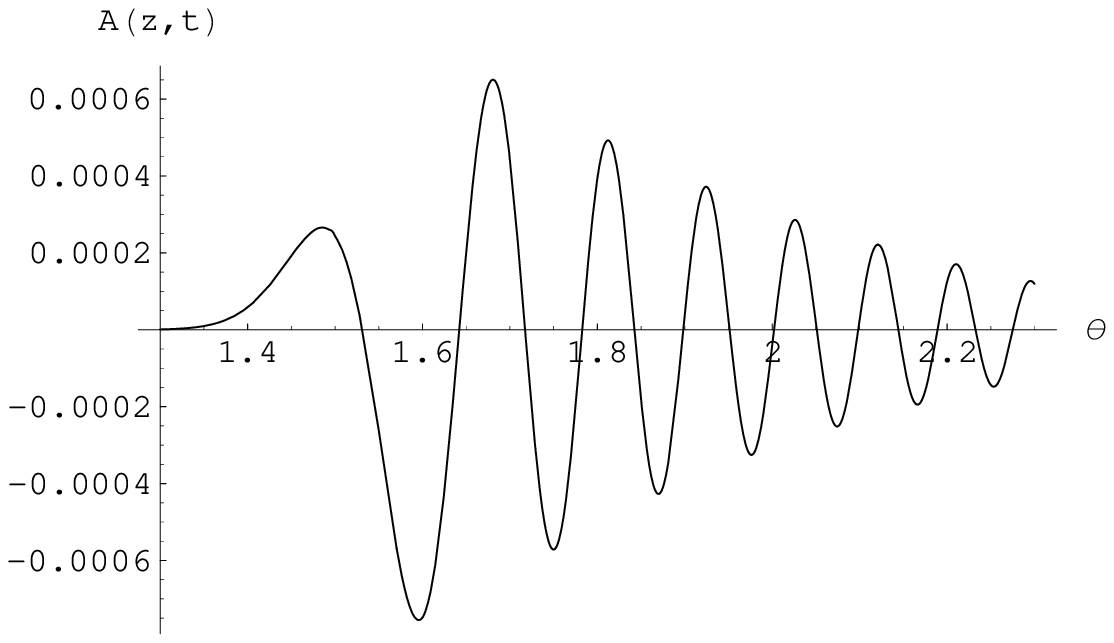}
\includegraphics[width=.7\textwidth]{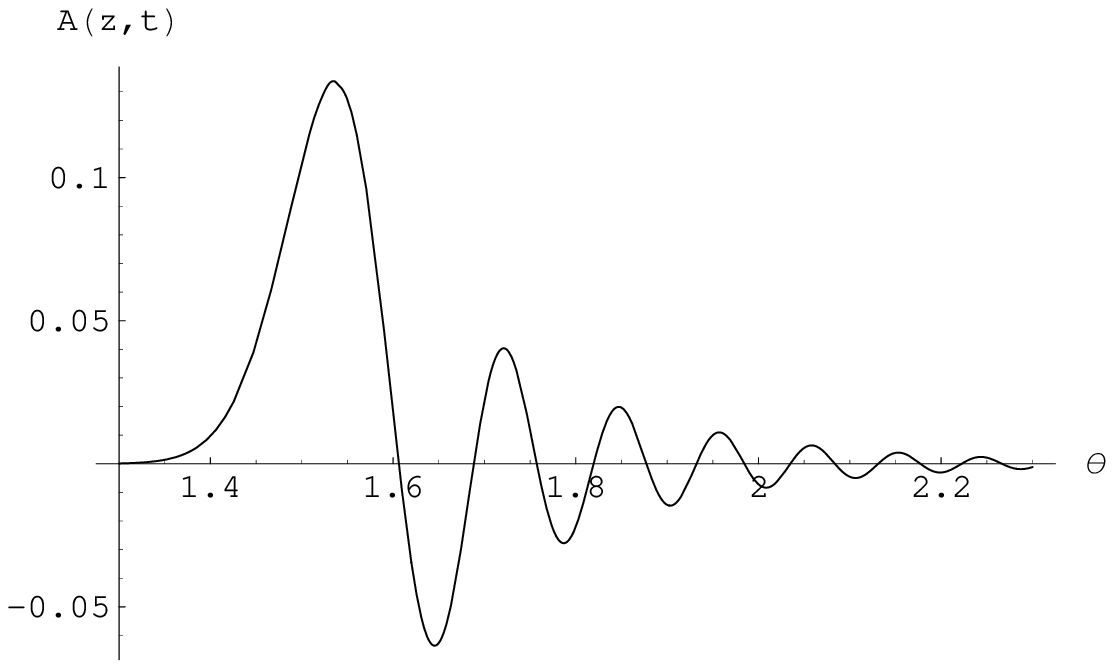}
\caption{\textit{Dynamic behavior of the Brillouin precursor in
the Lorentz medium described by Brillouin's parameters obtained
for: (top) $\b=2.0\times 10^{14} s^{-1}$, and (bottom)
$\b=2.0\times 10^{17} s^{-1}$. Here, $\o_c=2.5\times 10^{16}
s^{-1}$ and $\la=3.0\times 10^{-15} s$.}}
\end{figure}
An important question arises on how the rate parameter $\b$
affects the form of the Brillouin precursor.

As the parameter $\b$ in (\ref{e30}) increases starting from
relatively small values, the shape of the precursor remains
virtually unchanged while its magnitude grows. This tendency is no
longer valid if $\b$ enters a transitory interval. In that
interval the shape of the precursor changes and its magnitude
rapidly increases. Above transitory interval, further increase of
$\b$ leaves the shape and the magnitude of the precursor virtually
constant. The form of Brillouin precursor for $\b$ below
($2.0\times 10^{15} s^{-1}$) and above ($2.0\times 10^{17}
s^{-1}$) the transitory interval is shown in Fig.~3.

Explanation of this behavior lies in the properties of the
coefficients $c_0$ and $c_1$ in (\ref{e30}), which are
$\b$-dependent. The coefficients, in turn, determine the weight
with which Airy function and its derivative contribute to the
precursor.

\begin{figure}[h]
\centering
\includegraphics[width=.7\textwidth]{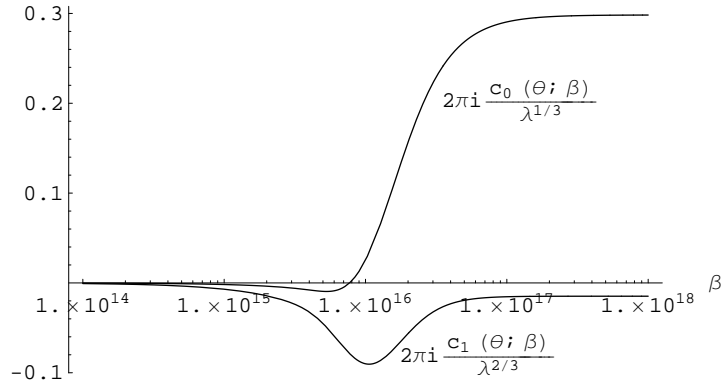}
\caption{\textit{Plots of $2\pi i\la^{-1/3} c_0(\t;\b)$ and $2\pi
i\la^{-2/3} c_1(\t;\b)$ against the speed parameter $\b$ at
$\t=1.502$. Here, $\o_c=2.5\times 10^{16} s^{-1}$ and
$\la=3.0\times 10^{-15} s$.}}
\end{figure}

First, consider the case of $\t<\t_s$. In Fig.~4 the coefficients
of, respectively, $\hbox{Ai}$ and $\hbox{Ai}^{\prime}$ in the
parentheses in (\ref{e30}) multiplied by $2\pi i$ are plotted
against $\b$. The value of $\t$ is chosen to be slightly below
$\t_s$. For relatively small $\b$ the term proportional to
$\hbox{Ai}^{\prime}(\cdot)$ dominates over the term proportional
to $\hbox{Ai}(\cdot)$. In this case the ratio of both terms
remains unchanged in a wide interval of $\b$ variation. The
magnitude of the precursor increases with $\b$ growth up to the
moment where the contribution from $\hbox{Ai}(\cdot)$ changes sign
and rapidly grows until finally it settles down at a virtually
constant level. At the same time the contribution from
$\hbox{Ai}^{\prime}(\cdot)$ decreases to another constant level
and is very small compared to the other term. At this stage the
shape and the magnitude of the precursor are approximately
determined by the special form of the function
\begin{equation}\label{e37}
g(\o)=
\frac{1}{4\pi}\left(\frac{1}{\o+\o_c}-\frac{1}{\o-\o_c}\right)
\end{equation}
which appears in $c_0$ and $c_1$, which is a limiting case of
$g(\o)$ as $\b \rightarrow \infty$, i.e.\ for the initial signal
with a unit step function envelope.

Now consider the case of $\t>\t_s$ where the precursor becomes
oscillatory. The envelope of the oscillations can be conveniently
approximated with the help of (\ref{e36}) by
\begin{equation}\label{e38}
\tilde{A}(z,t;\b)\approx 2
 e^{\la \mbox{\scriptsize{Re}}[\phi(\o_2)]}\left|
 \left(\frac{-2\pi}{\la \phi^{''}(\o_{2})}\right)^{1/2}
 g(\o_2;\b)\right|,
\end{equation}
provided $|\g(\t)|$ is sufficiently large.

\begin{figure}[h]
\centering
\includegraphics[width=.7\textwidth]{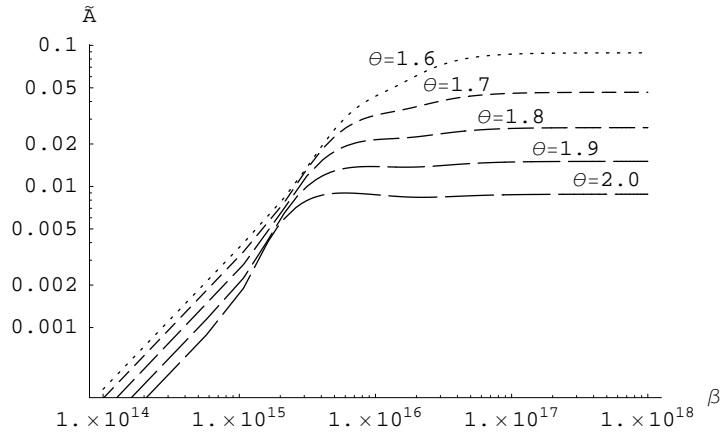}
\caption{\textit{Dependence of the magnitude of the Brillouin
precursor's envelope on $\b$. Calculated at different values of
$\t$. Here, $\o_c=2.5\times 10^{16} s^{-1}$ and $\la=3.0\times
10^{-15}$.}}
\end{figure}

In Fig.~5 the magnitude of the precursor envelope is plotted
against the parameter $\b$ for different values of $\t$. It is
seen again that after fast growth of the envelope magnitude at
relatively small values of $\b$, which occurs with approximately
the same rate for all $\t$, the magnitude reaches a saturation
level at higher values of $\b$. Since the saturation appears
earlier at larger values of $\t$, the precursor envelope has a
tendency to become narrower with growing $\b$. Additionally, one
observes that with growing $\b$ the first extremum moves towards
larger values of $\t$. This is a direct consequence of the fact
that the first extremum of the Airy function is shifted towards
negative values of its argument as compared to the first extremum
of the derivative of the Airy function. It has an additional
effect on narrowing the precursor shape.

\section{Conclusions}
In this paper we have derived the uniform and non-uniform
asymptotic representations for the Brillouin precursor in a
Lorentz medium, excited by an incident signal of finite rise time,
and well defined, startup time. With the use of these
representations we analyzed the effect of the speed parameter $\b$
on the form and magnitude of the precursor. The results obtained
can be helpful e.g.\ in applications involving triggering devices
that work with signal amplitudes close to the noise level. In this
paper we did not consider the problem of smooth transition from
Brillouin precursor to the main signal.

\vspace{1.5ex} \noindent \textbf{Acknowledgment}

\noindent The research presented in this work was partially
supported by the State Committee for Scientific Research under
grant 8 T11D 020 18.


\begin{thebibliography}{aa}

\bibitem{so;14} A.~Sommerfeld: \"{U}ber die Fortpflanzung des Lichtes
in disperdierenden Medien. Ann.\ Phys., Lepzig, 1914, vol.~{\bf
44}, pp.~177-202

\bibitem{br;14} L.~Brillouin: \"{U}ber die Fortpflanzung des Lichtes
in disperdierenden Medien. Ann.\ Phys., Lepzig, 1914, vol.~{\bf
44}, pp.~203-240

\bibitem{br;60} L.~Brillouin: {\it Wave Propagation and Group
Velocity}. New York, Academic, 1960

\bibitem{ks;96} M.~Kelbert and I.~Sazonov: \textit{Pulses and Other
wave Processes in Fluids}. Kluwer, 1996

\bibitem{os;97} K.~E.~Oughstun and G.~C.~Sherman: {\it Electromagnetic
Pulse Propagation in Causal Dielectrics}. vol.~\textbf{16},
Berlin, Springer, 1997

\bibitem{ac;97} A.~Ciarkowski: Asymptotic analysis of
propagation of a signal with finite rise-time in a dispersive,
lossy medium. Arch.~Mech., 1997, vol.~\textbf{49}, pp.~877-892

\bibitem{rg;51} I.~M.~Rhyzhik and I.~S.~Gradshteyn: {\it Tables of
Integrals, Sums, Series and Products}. 3-rd ed., National
Publishers of the Technical Literature, Moscow, 1951, Sec.~6.39
(in Russian)

\bibitem{cf;57} C.~Chester, B.~Friedman and F.~Ursell: An
extension of the method of steepest descents. Proc.\ of Cambridge
Phil.\ Soc., 1957, vol.~\textbf{53}, pp.~599-611

\bibitem{fm;73} L.~B.~Felsen and N.~Marcuvitz: {\it
Radiation and Scattering of Waves}. Prentice Hall, 1973, Ch.\ 4

\bibitem{bh;75} N.~Bleistein and R.~A.~Handelsman: {\it Asymptotic
Expansions of Integrals}. Holt, Rinehart and Winston, 1975, Ch.\ 9

\bibitem{ac;02} A.~Ciarkowski: Approximate representation for the
Brillouin precursor in a Lorentz medium. To appear in Electronics
and Telecommunication Quarterly, 2002, vol.~\textbf{48}

\bibitem{as;64} M.~Abramovitz and I.~A.~Stegun, Editors: {\it
Handbook of Mathematical Functions}, Nat.~Bureau of Standards,
1964, Sec.~10.4


\end{thebibliography}
\end{document}